\documentclass[superscriptaddress,twocolumn,prl,showpacs,preprintnumbers,amsmath,amssymb]{revtex4}

\usepackage[T1]{fontenc}\usepackage[latin1]{inputenc}
\usepackage{dcolumn,graphicx,color,booktabs,microtype}
\usepackage[charter]{mathdesign}

\begin{document}

\title{Highly anisotropic anomaly in the dispersion of the copper-oxygen bond-bending phonon in superconducting YBa$_2$Cu$_3$O$_{7}$ from inelastic neutron scattering}

\author{M. Raichle}
\affiliation{Max Planck Institute for Solid State Research, D-70569 Stuttgart, Germany}

\author{D. Reznik}
\affiliation{Physics Department, University of Colorado-Boulder,
Boulder, CO 80309, USA}

\author{D. Lamago}
\affiliation{Karlsruhe Institute of Technology, Institute for Solid State Physics, D-76021 Karlsruhe, Germany}
\affiliation{Laboratoire L\'{e}on Brillouin, CEA-CNRS, CE-Saclay, 91191 Gif-sur-Yvette, France}

\author{R. Heid}
\affiliation{Karlsruhe Institute of Technology, Institute for Solid State Physics, D-76021 Karlsruhe, Germany}

\author{Y. Li}
\affiliation{Max Planck Institute for Solid State Research, D-70569 Stuttgart, Germany}

\author{M. Bakr}
\affiliation{Max Planck Institute for Solid State Research, D-70569 Stuttgart, Germany}

\author{C. Ulrich}
\affiliation{Max Planck Institute for Solid State Research, D-70569 Stuttgart, Germany}
\affiliation{Australian Nuclear Science and Technology Organization (ANSTO), New South Wales 2234, Australia}
\affiliation{University of New South Wales, Sydney, New South Wales 2052, Australia}

\author{V. Hinkov}
\affiliation{Max Planck Institute for Solid State Research, D-70569 Stuttgart, Germany}

\author{K. Hradil}
\affiliation{Forschungsneutronenquelle Heinz Maier-Leibnitz (FRM-II), TU München, D-85747 Garching, Germany}

\author{C.T. Lin}
\affiliation{Max Planck Institute for Solid State Research, D-70569 Stuttgart, Germany}

\author{B. Keimer}
\affiliation{Max Planck Institute for Solid State Research, D-70569 Stuttgart, Germany}

\date{\today}

\begin{abstract}
Motivated by predictions of a substantial contribution of the
``buckling'' vibration of the CuO$_2$ layers to $d$-wave
superconductivity in the cuprates, we have performed an inelastic
neutron scattering study of this phonon in an array of untwinned
crystals of YBa$_2$Cu$_3$O$_{7}$. The data reveal a pronounced
softening of the phonon at the in-plane wave vector ${\bf q} =
(0, 0.3)$
upon cooling below $\sim 105$ K, but no corresponding anomaly at ${\bf
q} = (0.3, 0)$. Based on the observed in-plane anisotropy, we
argue that the electron-phonon interaction responsible for this
anomaly supports an electronic instability associated with a
uniaxial charge-density modulation and does not mediate $d$-wave
superconductivity.
\end{abstract}

\maketitle

Research on the mechanism of high-temperature superconductivity
in the cuprates has recently made substantial progress based on a
quantitative analysis of possible Cooper pairing interactions,
which include coupling to spin fluctuations and phonons. The spin
fluctuation mediated pairing interaction favors the
experimentally observed $d$-wave pairing state, \cite{abanov} but
its strength relative to phonon-mediated pairing interactions
remains a matter of intense debate.
\cite{abanov,gunnarsson,dicarlo,maksimov,savrasov,jepsen,johnston,sakai,honerkamp,giustino,heid,ishihara} Whereas density
functional theory (DFT) indicates a small \cite{savrasov,jepsen}
or negligible \cite{giustino,heid} contribution of the
electron-phonon interaction (EPI) to $d$-wave superconductivity, it has
been argued that electron correlations can greatly enhance the
coupling strength \cite{gunnarsson,dicarlo} such that it yields a
substantial \cite{johnston} or even dominant \cite{maksimov}
contribution to the pairing interaction. In this context, much
attention has been focused on phonons that modulate the Cu-O
bonds in the CuO$_2$ layers, which are generic to all cuprate
superconductors. Most model calculations indicate that the
EPI of high-energy stretching vibrations
of these bonds is either detrimental \cite{sakai,honerkamp} or
indifferent \cite{savrasov,jepsen,johnston} to $d$-wave pairing.
However, several theories indicate that coupling to the
lower-energy ``buckling'' vibration that modulates the Cu-O-Cu
bond angle is attractive in the $d$-wave channel.
\cite{savrasov,jepsen,johnston,sakai,honerkamp}

Two experimental strategies have been employed to test these
theories and to quantify the strength of the different
electron-boson coupling channels. The first approach has
identified fingerprints of coupling to bosonic modes in the
electronic spectral functions extracted from angle-resolved
photoemission (ARPES), optical, and tunneling data,
\cite{gunnarsson,abanov} but these data are not specific enough
to unequivocally discriminate between contributions from spin
excitations and phonons. A second, complementary approach
therefore targets corresponding anomalies in the bosonic
excitations. The prominent spectral-weight
redistribution of spin excitations at the superconducting
transition temperature determined by neutron scattering
\cite{hinkov} indicates a positive contribution to the $d$-wave
pairing susceptibility, in agreement with theoretical predictions.
\cite{abanov} The Cu-O bond-stretching vibration also exhibits
anomalies indicative of a substantial EPI,
\cite{chung,pintschovius,stercel,reznik_prb,uchiyama,reznik_nature,graf}
but most model calculations agree that coupling to this mode does not
support $d$-wave pairing. These, however, have not predicted or reproduced
the above-mentioned anomalies \cite{reznik_comment} (with the exception of Ref. \onlinecite{ishihara}).

\begin{figure*}[t]
\includegraphics[width=1.9\columnwidth,angle=0]{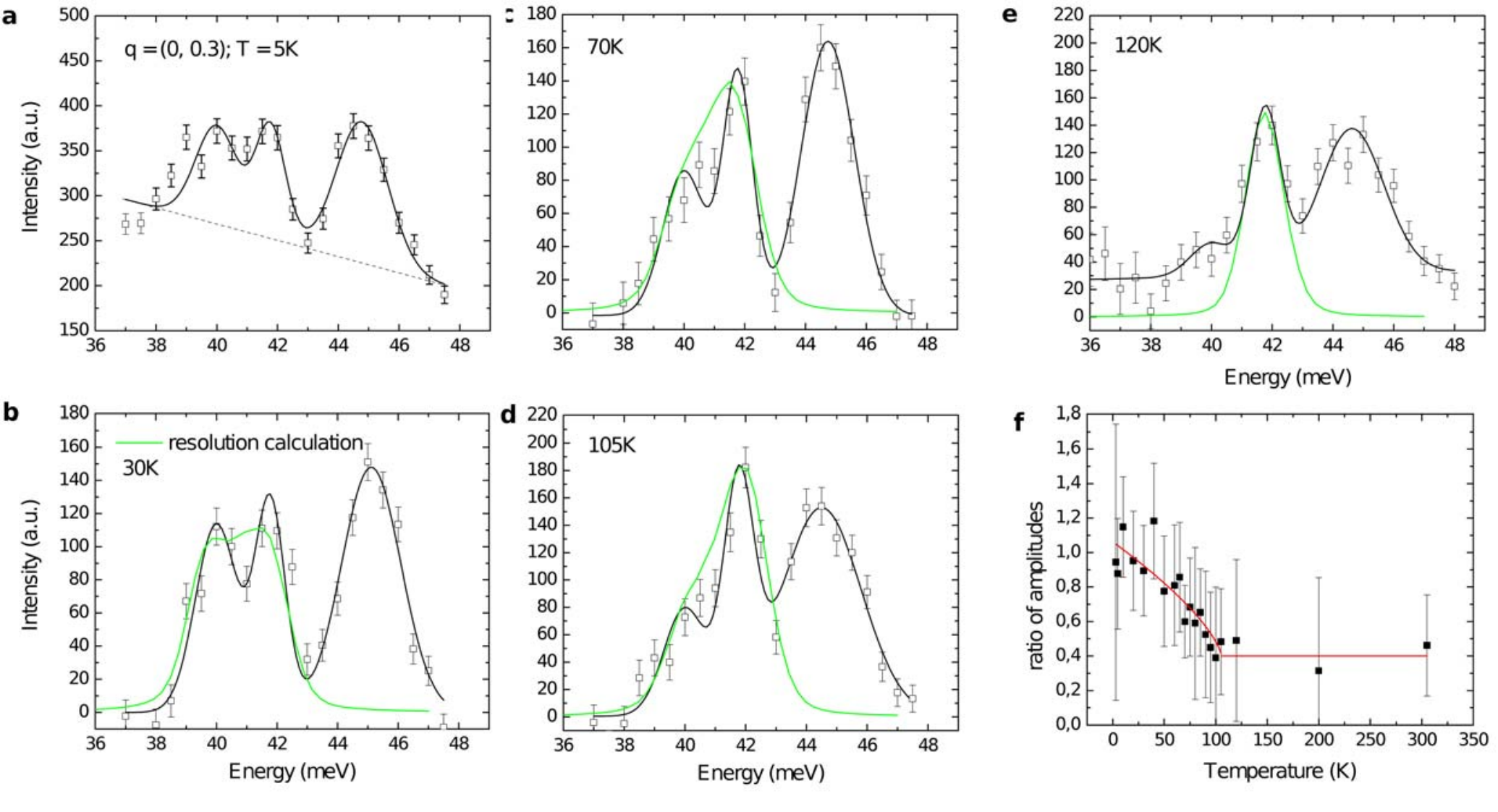}
\caption {(a-e) Constant-{\bf Q} scans at ${\bf Q} = (0,0.3,10)$.
The profile at energy 40-42 meV arises from the bucking mode.
Panel (a) shows raw data. A sloping background (dotted line) was
subtracted in the other profiles for clarity. The black and green
lines are results of fits to a phenomenological expression
comprising two Voigt functions and to a full resolution calculation,
respectively, as described in the text. (f) Temperature
dependence of the lineshape parameterized by the intensity of the
two peaks in the phenomenological expression.}
\end{figure*}

Much less information is available on the buckling mode, which
has been predicted to favor $d$-wave superconductivity. \cite{savrasov,jepsen,johnston,sakai,honerkamp} Raman
scattering experiments on optimally doped cuprates have shown
that the mode exhibits a superconductivity-induced softening of
$\sim 1.5$ \% at wave vector ${\bf q} =0$. \cite{friedl,bakr}
However, {\bf q}-dependent data are required for a comprehensive
assessment of the contribution of the buckling phonon to the
experimentally observed electron self-energy anomalies and to the
$d$-wave pairing state. Such data can be collected
by inelastic neutron scattering. Since the lattice dynamics is
further complicated by structural disorder,
prior neutron scattering work has focused on stoichiometric
YBa$_2$Cu$_3$O$_{7}$. \cite{pyka,reznik_prl} The results
confirmed a modest softening of the buckling mode upon cooling
below the superconducting transition temperature $T_c$, which
agrees with Raman scattering data \cite{friedl,bakr} at ${\bf q}
= 0$ and decreases rapidly with increasing {\bf q}, in accord with
model calculations. \cite{fukuyama,devereaux_prb} However, a
broadening of the neutron profiles with a maximum amplitude at
nonzero ${\bf q}$ \cite{reznik_prl} was not explained by these
calculations. Since these experiments were carried out on twinned
specimens, the in-plane anisotropy of these features could not be
resolved.

In the light of recent reports of pronounced, strongly
temperature and doping dependent in-plane anisotropies of the
magnetic \cite{hinkov} and transport \cite{ando,daou} properties
of underdoped YBa$_2$Cu$_3$O$_{6+\delta}$, we have used inelastic
nuclear neutron scattering to determine the in-plane anisotropy
of the dispersion of the buckling mode in an untwinned specimen of
YBa$_2$Cu$_3$O$_{7}$. Our sample consists of 185 individually
detwinned single crystals with total mass 2.63 g, co-aligned on an
aluminum plate. Magnetization measurements on a representative
set of crystals showed that $T_c = 90.0 \pm 0.5$ K. X-ray powder
diffraction measurements yielded the room temperature lattice
parameters $a=3.8173$, $b=3.8844$, and $c = 11.681$ \AA. These
data imply that the sample is nearly stoichiometric ($\delta \sim 1$)
and slightly overdoped. This is important because even a
small density of oxygen vacancies leads to the formation of
superstructures that may affect the phonon dispersions.
\cite{strempfer} The detwinning ratio of the array is 86
percent. The neutron measurements were performed on the PUMA
spectrometer at the FRM-II reactor in Garching, Germany, and on
the 1T spectrometer at the ORPHEE reactor at LLB, Saclay, France.
The (2,2,0) reflection of Cu and the (0,0,2) reflection of
pyrolytic graphite were used to monochromate and analyze the
neutron beam, respectively. The final neutron energy was fixed at
8 meV, and the data were taken in the Brillouin zones adjacent to
the (1,0,10), (0,1,10), and (0,3,1) reciprocal lattice vectors.
(Wave vectors are quoted in units of $2\pi / (a,b,c)$.)

Figure 1 shows representative neutron scattering profiles in a
scattering geometry that maximizes the structure factor of the
buckling mode, at the in-plane wave vector ${\bf q} = (0, 0.3)$
where the broadening was observed in the early experiment
\cite{reznik_prl}. In addition to the buckling mode at energy $E
= 40-42$ meV, the data show a mode at 44.5 meV that is not
strongly temperature dependent and will not be discussed further.
The profiles of the buckling mode, on the other hand, exhibit a
highly unusual behavior. At temperature $T = 120$ K, the profile
shows a single peak with a barely discernible shoulder on the
low-energy side (Fig. 1e). Upon cooling, however, the shoulder
grows continuously and draws intensity from the main peak,
resulting in a two-wing profile at low temperatures (Figs. 1a-d).
In order to parameterize the temperature evolution of the
lineshape, we have fitted the data to a phenomenological form
that comprises two Voigt functions with fixed energies 40.0 and 41.8
meV, and widths 1.6 and 1.2 meV (black lines in Figs. 1a-e).
The fits yield a good description of the experimental data over the
entire temperature range (although they do not correctly capture
the physics; see below).
The $T$-dependence of the
resulting peak intensities (Fig. 1f) indicates a well defined
onset of the lineshape anomaly at $T = 105 \pm 15$ K.

This unusual behavior is confined to in-plane wave vectors in the
range ${\bf q} = (0, 0.2) \rightarrow (0, 0.4)$. The profiles for
wave vectors outside this range along the same direction, and for
the entire range along the perpendicular direction, ${\bf q} =
(0, 0) \rightarrow (0.5, 0)$, are well described by a single,
undistorted peak. A representative scan at ${\bf q} = (0.3, 0)$
is shown in Fig. 2. The strong in-plane anisotropy of the
anomalous dispersion of the buckling mode revealed by these data
could not be recognized in previous work on twinned crystals.
\cite{pyka,reznik_prl}

\begin{figure}[t]
\includegraphics[width=0.9\columnwidth,angle=0]{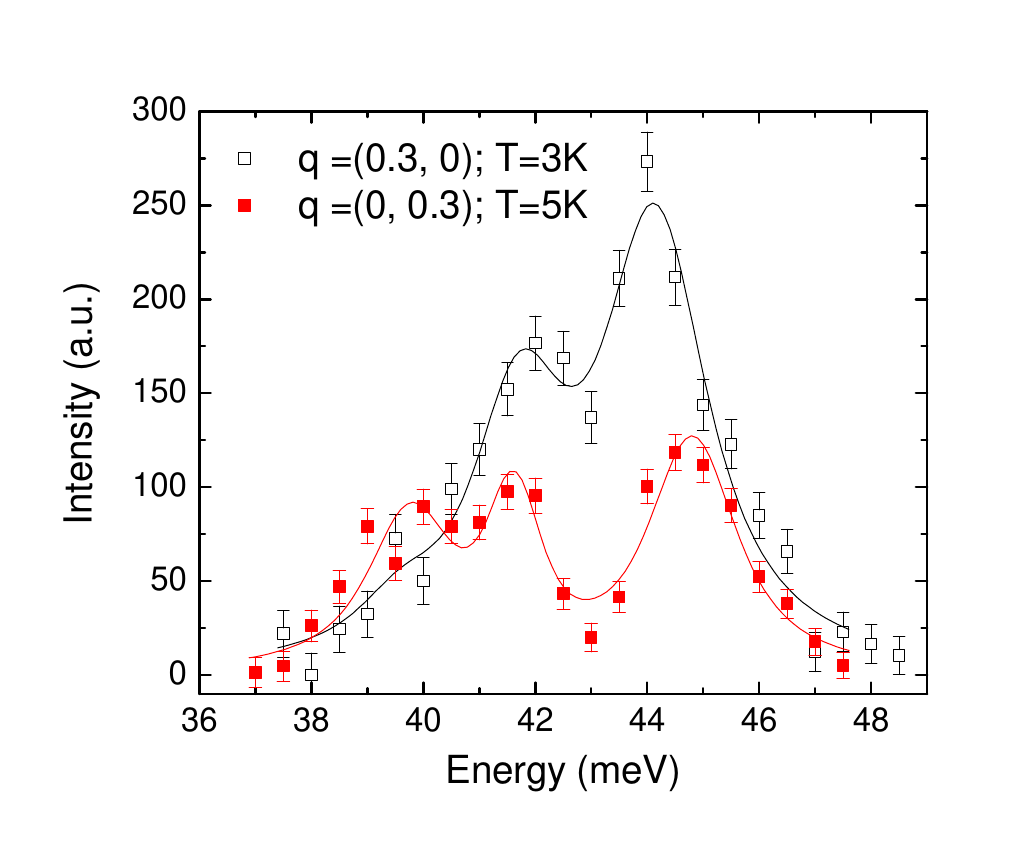}
\caption {Constant-{\bf Q} scans of the buckling mode at ${\bf Q}
= (0.3,0,10)$ and $(0,0.3,10)$. A sloping background (Fig. 1a) was
subtracted.}
\end{figure}

Since the $E-{\bf q}$ range of the anomalous low-temperature
dispersion of the buckling mode is comparable to the dimensions of
the resolution ellipsoid of the neutron spectrometer, a careful
resolution calculation is required to extract information from
the observed lineshape. We have performed such a calculation
based on the dispersion relation shown in Fig. 3, which is flat
along the $a$-direction at all temperatures (Fig. 3a) and along
$b$ for $T \geq 120$ K (red line in Fig. 3b), but develops a
sharp dispersion anomaly along $b$ at lower temperatures (Fig. 3b). Considering the simplicity of the model, the agreement between the
calculated profiles (green lines in Figs. 1
b-e) and the data is satisfactory. In
particular, the calculation provides an explanation of the two-wing
profile discussed above. The low-energy wing results from the
bottom of the dispersion anomaly at the nominal spectrometer
setting. However, the instrumental resolution ellipsoid also
encompasses a wide segment of the nearly flat dispersion surface
away from the anomaly.
This explains the high-energy wing of the profile.

\begin{figure}[t]
\includegraphics[width=0.9\columnwidth,angle=0]{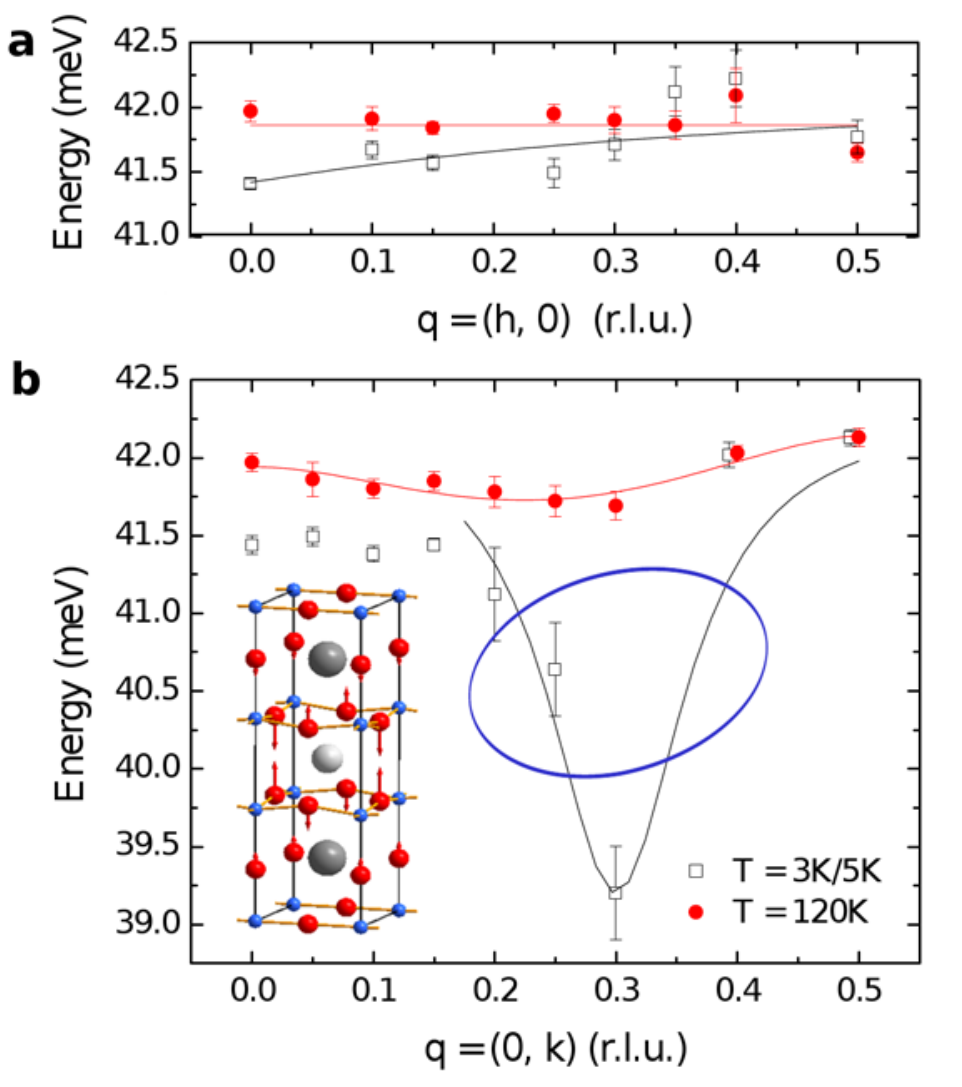}
\caption {Dispersion of the buckling mode along the $a$- and
$b$-axes. The black line is the dispersion used for the resolution
calculation. A projection of the four-dimensional resolution
ellipsoid is shown for comparison. The data point at ${\bf q} = (0, 0.3)$
is the result of the resolution convolution described in the text,
the remaining points were determined by fits to standard Voigt functions. The inset shows the
eigenvector of the buckling mode at ${\bf q} = (0, 0.3)$. The
elongations of the apical oxygen atoms and of the in-plane oxygen
atoms along $b$ were enlarged by a factor of four for clarity.}
\end{figure}


A highly anisotropic dispersion anomaly of the buckling mode that
is strongly enhanced upon cooling below $\sim 105$ K thus
provides an adequate explanation of all aspects of the
experimental data. In view of the presence of several other
phonon branches in the energy range 35-45 meV previously detected
by Raman and infrared spectroscopies \cite{humlicek}, one may
wonder to what extent the $T$-dependent spectral-weight
redistribution of the buckling mode results from mixing with other
modes. We have calculated the neutron structure factors based on
the eigenvectors provided by ab-initio DFT
calculations \cite{bohnen} and found that most of the modes in
this energy range (mostly in-plane vibrations of the apical
oxygen ions, with some admixture of chain and plane oxygens)
exhibit substantial spectral weight in the Brillouin zone around
the reciprocal lattice vector $(0, 3, 1)$. Scans in this zone for
${\bf q} = (0, 0.3)$ show no evidence of temperature dependent
modifications of the phonon intensities or lineshapes outside the
statistical error. While a spectral-weight transfer to
other modes
analogous to the one discussed for the Cu-O
stretching vibration \cite{pintschovius,reznik_prb}
cannot be ruled out entirely, it therefore does not appear to play a major
role for the buckling mode.
This conclusion is also supported by
the observation that the integrated intensity of the buckling
mode profile shown in Fig. 1 is $T$-independent.

We now discuss the microscopic origin of the dispersion anomaly.
Since wave vectors around ${\bf q} \sim (0, 0.3)$ connect nested
segments of the Fermi surface of YBa$_2$Cu$_3$O$_{7}$,
\cite{andersen} an anomaly resulting from the EPI provides a natural explanation of our data.
The strong temperature dependence of the anomaly (Fig. 1) can
then be attributed to the opening of a gap (or pseudogap) on the Fermi surface. Although the
data of Fig. 1f are consistent with a sharp onset or
enhancement at $T_c$, the momentum-space structure of the anomaly speaks
against a direct relation to the superconducting gap.
\cite{bakr,nemetschek,kirtley}
Raman scattering experiments have shown that the pair-breaking
energy $2 \Delta \approx 60$ meV, with a $\sim 10$ \% anisotropy
between $a$- and $b$-directions. \cite{bakr,nemetschek} Since
$2\Delta$ significantly exceeds the phonon energy along both $a$-
and $b$-axes, the onset of superconductivity is expected to
affect the phonon lineshapes similarly in both directions. Specifically,
a softening of the phonon is expected as the gap sweeps through the phonon energy upon cooling.
\cite{allen,weber} This is indeed the case for the
superconductivity-induced softening centered at ${\bf q} =0$
(Fig. 3), \cite{friedl,bakr} but not for the anomaly at ${\bf q} \neq 0$, which is
present only along the $b$-axis. The strong anisotropy in phonon softening is indicative
of a uniaxial EPI, which cannot be
responsible for the superconducting gap whose magnitude is similar
along $a$ and $b$, as further explained below.

Based on these considerations, a (real or incipient) instability other than
superconductivity must be responsible for the
$T$-dependent anomaly of the buckling mode. The strong in-plane
anisotropy of the anomaly indicates that it has a one-dimensional
structure, which may arise from a uniaxial charge density wave
(CDW) instability. There is independent evidence for a CDW
originating in the CuO chains in the YBa$_2$Cu$_3$O$_{7}$ crystal
structure, \cite{mehring,berthier,liu} and the wave vector of the
phonon anomaly is similar to the one of the CDW recently reported
by x-ray diffraction. \cite{liu} The temperature onset of the
phonon anomaly (Fig. 1f) is also near the onset of a pronounced
in-plane anisotropy of the Nernst effect in YBa$_2$Cu$_3$O$_{7}$,
\cite{daou} which has been interpreted as evidence of a stripe or
Pomeranchuk instability originating in the CuO$_2$ planes. We
note, however, that the wave vector of the instability indicated
by the phonon anomaly is perpendicular to the one characterizing
the uniaxial spin density wave state recently observed in
strongly underdoped YBa$_2$Cu$_3$O$_{6+\delta}$. \cite{haug} This
may imply that spin (charge) driven instabilities with
propagation vector along $a$ ($b$) dominate the interplay with
$d$-wave superconductivity in the low (high) doping regimes of
the YBa$_2$Cu$_3$O$_{6+\delta}$ phase diagram. Such a multi-phase
competition may be influenced by the hybridization between
the chain-derived energy bands with those arising from
the CuO$_2$ layers. \cite{andersen}

We finally address the magnitude of the dispersion anomaly. If we assume that the phonon softens due to an opening of a
gap on the Fermi surface, (see Ref. \onlinecite{allen} for the case of the superconducting gap, although our argument would
work for a gap of any origin) the phonon renormalization magnitude
should be proportional to the {\bf q}-dependent EPI strength. The
observed $\sim 6$ \% softening of the buckling phonon at ${\bf q} \sim (0, 0.3)$ upon
cooling indicates that the EPI of the
buckling mode may be strong enough to contribute significantly to
the electronic self-energy along the antinodal direction of the
$d$-wave gap function, as hypothesized based on ARPES
measurements. \cite{gunnarsson,johnston} Our data provide an
improved basis for a quantitative assessment of this contribution
relative to the one generated by spin fluctuations
\cite{abanov,hinkov}. Despite its strength,
we have shown that the highly anisotropic EPI responsible for the anomaly at ${\bf q} \sim (0, 0.3)$ does not support $d$-wave superconductivity.
We conclude that the buckling mode does not contribute to
$d$-wave pairing beyond ${\bf q} =0$, where the phonon softening below $T_c$ translates
into a small dimensionless coupling constant $\lambda_d \sim
0.02$, in agreement with DFT calculations. \cite{heid} It is possible, however, that it
contributes to the observed sub-dominant $s$-wave admixture to
the pairing state. \cite{bakr,nemetschek,kirtley}

In any case, the sharp phonon anomaly indicates that overdoped YBa$_2$Cu$_3$O$_{7}$ is close to an instability associated with a uniaxial CDW.
Further work is required to assess the contribution of the electron-phonon coupling driving this instability to the anomalous band dispersions revealed by ARPES.



We thank O.K. Andersen, C. Bernhard, P. Bourges, M. Cardona, V. Damljanovic, O. Gunnarsson, T. Keller, G. Khaliullin, D. Manske, L.
Pintschovius, Y. Sidis, and R. Zeyher for discussions, and R.E. Dinnebier and D. Haug for help with the x-ray
measurements.

\end{document}